 \documentclass[final,3p,times,twocolumn]{elsarticle}
\usepackage{amssymb}
\usepackage{graphicx}
\usepackage{geometry}
\usepackage{xcolor}
\usepackage{amsmath}
\usepackage{amsfonts}
\usepackage[colorlinks,citecolor=red,urlcolor=red]{hyperref}
\usepackage{amsthm}

\newcommand{\ket}[1]{|#1\rangle}
\newcommand{\bra}[1]{\langle #1|}

\newcommand{\prom}[1]{\langle #1\rangle}

\usepackage{color}
\definecolor{Rojo}{rgb}{1.0,0.0,0.0}
\journal{Computer Physics Communications}
\begin{document}
\begin{frontmatter}
%
\title{Green's functions technique for calculating the emission spectrum in a quantum dot-cavity system}
\author[author1]{Edgar A. G\'omez\corref{cor}}
\cortext[cor]{Corresponding author.\\\textit{E-mail address:} eagomez@uniquindio.edu.co}
\address[author1]{Programa de F\'isica, Universidad del Quind\'io, Armenia, Colombia}
\author[author2,author3]{J. D. Hern\'andez-Rivero}
\author[author2]{Herbert Vinck-Posada}
\address[author2]{Departamento de F\'isica, Universidad Nacional de Colombia, Bogot\'a, Colombia}
\address[author3]{Departamento de F\'isica, Universidade Federal de Minas Gerais, Belo Horizonte MG, Brazil}
\begin{abstract}
We introduce the Green's functions technique as an alternative theory to the quantum regression theorem formalism for calculating the two-time correlation functions in open quantum systems. In particular, we investigate the potential of this theoretical approach by its application to compute the emission spectrum of a dissipative system composed by a single quantum dot inside of a semiconductor cavity. We also describe a simple algorithm based on the Green's functions technique for calculating the emission spectrum of the quantum dot as well as of the cavity which can easily be implemented in any numerical linear algebra package. We find that the Green's functions technique demonstrates a better accuracy and efficiency in the calculation of the emission spectrum and it allows to overcome the inherent theoretical difficulties associated to the direct application of the quantum regression theorem approach.
\end{abstract}
\end{frontmatter}
\section{Introduction}
The measurement and control of light produced by quantum systems have been the focus of interest of cavity quantum electrodynamics \cite{Walther:2006,Kavokin:2007}. Specially, the emission of light powered by solid-state devices coupled to nanocavities is an extensive area of research due to its promising technological applications, such as infrared and low-threshold lasers \cite{Altug:2006,Mu:1992}, single and entangled photon sources \cite{Stevenson:2006,Stace:2003}, as well as various applications in quantum cryptography \cite{Gisin:2002}, and quantum information \cite{Monroe:2002}. Experiments with semiconductor quantum dots (QDs) embedded in microcavities have revealed a plethora of quantum effects and offer desirable properties for harnessing coherent quantum phenomena at the single photon level. For example, the Purcell enhancement \cite{Todorov:2007}, photon anti-bunching \cite{Wiersig:2009}, vacuum Rabi splitting \cite{Khitrova:2006} and strong light matter coupling \cite{Reithmaier:2004}. These and 
many others quantum phenomena are being confirmed experimentally by observing the power spectral density of the light (PSD) emitted by quantum-dot-cavity systems (QD-Cavity). Thus, the PSD or so-called emission spectrum is the only relevant information about the system which allows to study the properties of light via measurements on correlations functions as stated by the Wiener-Khintchine theorem \cite{Mandel:1997}. In order to compute the absorption or emission spectrum in open quantum systems, more precisely, in QD-Cavity systems different approaches have been developed from theoretical point of view. For example, the method of the thermodynamic Green functions which is applied to the determination of the susceptibilities and absorption spectra of atomic systems embbeded in nanocavities \cite{Jedrkiewicz:2000}, and the time-resolved photoluminescence approach whose application allows to determine the emission spectrum by consideration of an additional subsystem called the photon reservoir \cite{Hieu:2010}. However, these methods have their own approximations and restrictions and therefore are not widely used. Frequently, the emission spectrum in QD-Cavity systems is computed through the Quantum Regression Theorem (QRT) \cite{Walls:1994,Lax:1966,Swain:1981}, since it relates the evolution of mean values of observables and the two-time correlation functions. It is worth mentioning that this approach can be difficult to 
implement in a computer program, it due to that computational complexity of QRT approach increases significantly as the number of 
QDs or modes inside the cavity, and the dimensionalities of the Hilbert spaces are large. In general, this approach is time-consuming due to that it requires to solve a large system of coupled differential equations, and numerical instabilities can arise. Moreover, theoretical complications can appears related to dynamics of the operators involved, as we will point out in the next section. In spite of this, the QRT approach is widely used for theoretical works, for example, in studies of the luminescence spectra of coupled light-matter systems in microcavities in the presence of a continuous and incoherent pumping \cite{delValle:2009,Quesada:2011}, and the relation between dynamical regimes and entanglement in QD-Cavity systems \cite{Vera:2009,Ishida:2013}. In the past, the Green's functions technique (GFT) was successfully applied for calculation of the micromaser spectrum \cite{Quang:1993}, as a methodology in which the two-time correlation function is treated as a Green's function that decays as the off-diagonal elements of the reduced density matrix of the system for a very specific initial conditions. Nevertheless, this approach has not been widely noticed in many significant situations in open quantum systems. Possibly, it is due to their work having a limitation of implementation.  
The purpose of this work is to present a simple, but efficient numerical method based on QRT formalism which overcome the inherent difficulties associated to the direct application of the QRT, by solving the dynamics of the system in the frequency domain directly. This paper is organized as follows: Section \ref{sec:QRT} review the theoretical background of quantum regression theorem and its relationship with the Green's functions technique. Section \ref{sec:appl} deal with a concrete application of our proposed method for computing the emission spectra of a dissipative QD-Cavity system. In section \ref{results} we show the numerical calculations of the emission of spectrum for the cavity and the quantum dot from both GFT and QRT approaches. Finally, we conclude in the last section.
\section{Theoretical background}\label{sec:QRT}
One of the most important measurements when the light excites resonantly a QD-Cavity system is the emission spectrum of the system. From theoretical point of view, it is assumed that corresponds to a stationary and ergodic process which can be calculated as a PSD of light using the well-known Wiener-Khintchine theorem \cite{Mandel:1997}. It states that the emission spectrum is given by the Fourier Transform of the correlation function (two-time expectation value) of the operator field $\hat{c}$, 
\begin{equation}
S(\omega)=\frac{1}{\pi n_c}\mathfrak{Re}\lim_{t\to\infty}\int_{0}^\infty K(\tau)  e^{i\omega \tau} d\tau,
\end{equation}
where $K(\tau)=\langle \hat{c}^{\dagger}(t+\tau)\hat{c}(t)\rangle$ and the normalizing factor is the population $n_c$ at the steady-state. In order to calculate the two-time expectation value is frequently used the QRT which states that if a set of operators $\{\hat{O}_{j}(t+\tau)\}$ satisfy
the dynamical equations $\frac{d}{d\tau}\prom{\hat{O}_{i}(t+\tau)}=\sum_{j}L_{ij}\prom{\hat{O}_j(t+\tau)}$
then $\frac{d}{d\tau}\prom{\hat{O}_{i}(t+\tau)\hat{O}(t)}=\sum_{j}L_{ij}\prom{\hat{O}_j(t+\tau)\hat{O}(t)}$ is valid for any operator $\hat{O}(t)$ at arbitrary time $t$. It is worth mentioning that vality of this theorem holds whenever a closed set of operators are involved in the dynamics. In general, to obtain the closed set of operators can be difficult or an impossible task, since it must be added as many operators as necessary in order to close the dynamics of the system. For example, in order to compute the emission of spectrum in a simple model of QD-Cavity system \cite{Quesada:2011,Vera:2009} two new operadors are required due to that the field operators in the interaction picture does not lead to a complete set. Before we consider the Green's functions technique, we will briefly describe the calculation of the QRT in an alternate form which will be the starting point in the following section. Lets consider a system operador $\hat{A}$ which does not operate on the reservoir, then its single-time expectation value in the Heisenberg picture is given by  
\begin{equation}\label{sec:QRT:01}
\prom{\hat{A}(t+\tau)}=Tr_{S\otimes R}[\hat{A}(t+\tau)\hat{\rho}_{S\otimes R}(t)]. 
\end{equation}
The operator $\hat{\rho}_{S\otimes R}(t)=\hat{\rho}_{S}(t)\otimes\hat{\rho}_{R}(t)$ depics the composite density operador of the system and reservoir. It is worth pointing out that the dynamics of the system depends directly on $\hat{\rho}_{S\otimes R}(t)$ for all times, but the validity of the Markovian approximation requires that the state of the system is sufficiently well described by $\hat{\rho}_{S}(t)=Tr_R(\hat{\rho}_{S\otimes R}(t))$, therefore it is sufficient to write $\hat{\rho}_{S\otimes R}(t)=\hat{\rho}_{S}(t)\otimes\hat{\rho}_{R}(t)$. In what follows, we change to the Schr\"odinger representation using $\hat{A}(t+\tau)=\hat{U}^{\dagger}(t+\tau,t)\hat{A}(t)\hat{U}(t+\tau,t)$ with $\hat{U}(t+\tau,t)$ being the unitary time-evolution operator, and after tracing over degrees of freedom of the reservoirs, we have
\begin{equation}\label{sec:QRT:02}
\prom{\hat{A}(t+\tau)}=Tr_{S}[\hat{A}(t)\hat{\rho}_{S}(t+\tau)], 
\end{equation}
where the reduced density operador for the system is given by $\hat{\rho}_S(t+\tau)=Tr_{R}[\hat{U}(t+\tau,t)\hat{\rho}_{S\otimes R}(t)\hat{U}^{\dagger}(t+\tau,t)]$. Then, if the $\hat{\rho}_S(t+\tau)$ satisfies the Markovian master equation $d\hat{\rho}_S(t+\tau)/d\tau=\mathcal{L}\hat{\rho}_S(t+\tau)$ with $\mathcal{L}$ the Liouvillian superoperator, the evolution of $\prom{\hat{A}(t+\tau)}$ can be computed by solving the dynamics of the master equation. To calculate the two-time correlation function $\prom{\hat{A}(t+\tau)\hat{B}(t)}$ where $\hat{A}(t+\tau)$ and $\hat{B}(t)$ are arbitrary Heisenberg operators, we proceed in a similar manner, it is,
\begin{eqnarray}\label{sec:QRT:03}
\prom{\hat{A}(t+\tau)\hat{B}(t)}&=&Tr_{S\otimes R}[\hat{A}(t+\tau)\hat{B}(t)\hat{\rho}_{S\otimes R}(t)],\notag \\ 
&=&Tr_S[\hat{A}(t)\hat{G}(t+\tau)],
\end{eqnarray}
where we have used the well-known composition and inversion properties of the evolution operator. Then, the two-time operator is given by 
\begin{equation}\label{sec:QRT:04}
 \hat{G}(t+\tau)=Tr_{R}[\hat{U}(t+\tau,t)\hat{B}(t)\hat{\rho}_{S\otimes R}(t)\hat{U}^{\dagger}(t+\tau,t)].
\end{equation}
By comparison of the Eq.~(\ref{sec:QRT:02}) and Eq.~(\ref{sec:QRT:03}), we find that $\hat{G}(t+\tau)$ is an operator that obeys the same dynamical equations as $\hat{\rho}_S(t+\tau)$, but as function of $\tau$. It is, $d\hat{G}(t+\tau)/d\tau=\mathcal{L}\hat{G}(t+\tau)$ with the boundary condition $\hat{G}(t)=\hat{B}(t)\hat{\rho}_S(t)$ at arbitrary time $t$. Hence, in the long-time limit the QRT reads,
\begin{equation}\label{eq-final}
\lim_{t\to\infty}\prom{\hat{A}(t+\tau)\hat{B}(t)}=Tr_S[\hat{A}\hat{G}(\tau)]
\end{equation}
where $\hat{G}(\tau)=Tr_{R}[\hat{U}(\tau)\hat{B}\hat{\rho}^{(ss)}_{S\otimes R}\hat{U}^{\dagger}(\tau)]$ is the Green's functions operator, and the operators $\hat{A}$, $\hat{B}$ and $\hat{\rho}^{(ss)}_{S\otimes R}$ are written in the Schr\"odinger representation. The superscript "(ss)" refers to the steady state of the reduced density operator of the system. After taking the Laplace transform on Eq.~(\ref{eq-final}), we obtain an expression for the emission of spectrum in terms of the Green's functions operator, it is,
\begin{equation}
S(\omega)=\frac{1}{\pi n_c}\mathfrak{Re}\,Tr_S[\hat{A}\hat{\tilde{G}}(i\omega)].
\end{equation}
Prior to leaving this section, we mention that this equation will be considered for computing the emission spectrum due to the cavity as well as the quantum dot, e.g. by considering the photon and fermionic operators in a separated way. Therefore, we will describe in the next subsetion a general approach that can be applied for both cases.
\subsection{Algorithm for the Green's functions technique}
Before to describe a simple algorithm for calculating the emission spectrum, we take into account that the dynamics for both opertors $\hat{G}(\tau)$ and $\hat{\rho}_S(\tau)$ are governed by the same Master equation, i.e., $d\hat{G}(\tau)/d\tau=\mathcal{L}\hat{G}(\tau)$ with $\mathcal{L}$ the Liouvillian superoperator, that efectivelly has a larger tensor rank than the reduced density operator of the system. So, we can write the dynamical equations for the Green's functions operator in a component form: 
\begin{equation}\label{components}
\frac{dG_{\tilde{\alpha}}(\tau)}{d\tau}=\sum_{\tilde{\beta}}\mathcal{L}_{\tilde{\alpha}\tilde{\beta}}G_{\tilde{\beta}}(\tau),
\end{equation}
together with the initial condition $G_{\tilde{\beta}}(0)$. The symbol $\tilde{\alpha}$ is a composite index for labeling the states of the reduced density operator of the system, e.g. for indexing both matter and photon states in the QD-Cavity system, see section \ref{sec:appl} for example. Hence, $G_{\tilde{\beta}}$ and $\mathcal{L}_{\tilde{\alpha}\tilde{\beta}}$ acts as a column vector and a matrix in this notation. (ii) To obtain the solution to the Eq.~(\ref{components}) in frequency domain via the Laplace Transform, it is $-\tilde{G}_{\tilde{\alpha}}(0)=\sum_{\tilde{\beta}}(\mathcal{L}_{\tilde{\alpha}\tilde{\beta}}-i\omega\delta_{\tilde{\alpha}\tilde{\beta}})\tilde{G}_{\tilde{\beta}}(i\omega),$ (iii) We perform the invertion of the matrix  
$\mathcal{M}_{\tilde{\alpha}\tilde{\beta}}=(i\omega\delta_{\tilde{\alpha}\tilde{\beta}}-\mathcal{L}_{\tilde{\alpha}\tilde{\beta}}$) and finally, the spectrum of emission is computed in terms of the initial conditions given by, 
\begin{equation}
\tilde{G}_{\tilde{\beta}}(i\omega)=\sum_{\tilde{\alpha}}\mathcal{M}^{-1}_{\tilde{\beta}\tilde{\alpha}}\tilde{G}_{\tilde{\alpha}}(0).
\end{equation}
These initial conditions are easily obtained by evaluating the Green's function operator at $\tau=0$. 
\section{Application to the quantum dot-cavity system}\label{sec:appl}
\subsection{Model}\label{sec:model}
\begin{figure}[h]
\centering
\includegraphics[scale=.19]{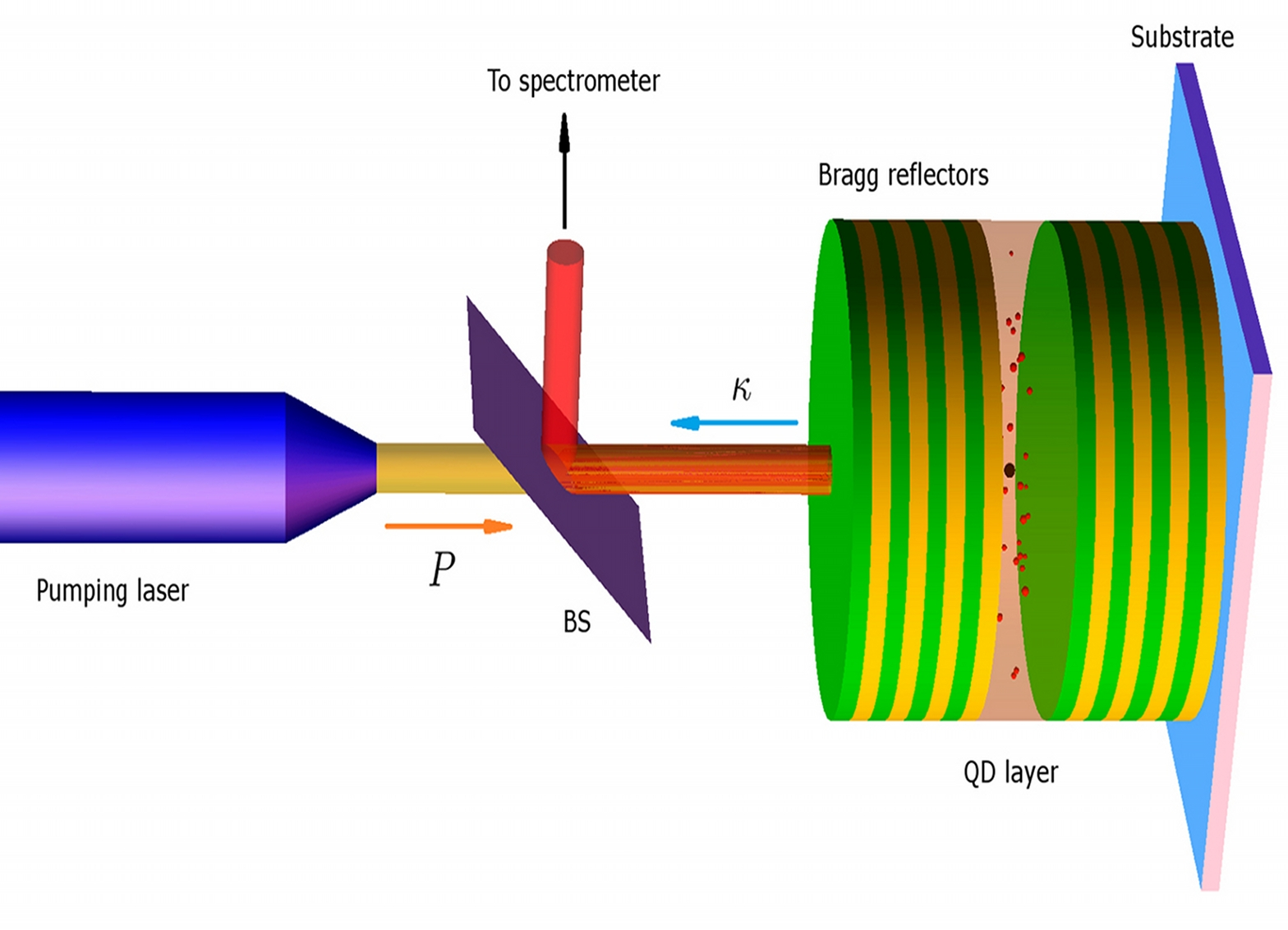}
\caption{Scheme of the simplified model of the QD-cavity system showing the processes of continuous pumping $P$ and cavity loses $\kappa$.}\label{systemscheme}
\end{figure}

\begin{figure}[h]
\centering
\includegraphics[scale=.5]{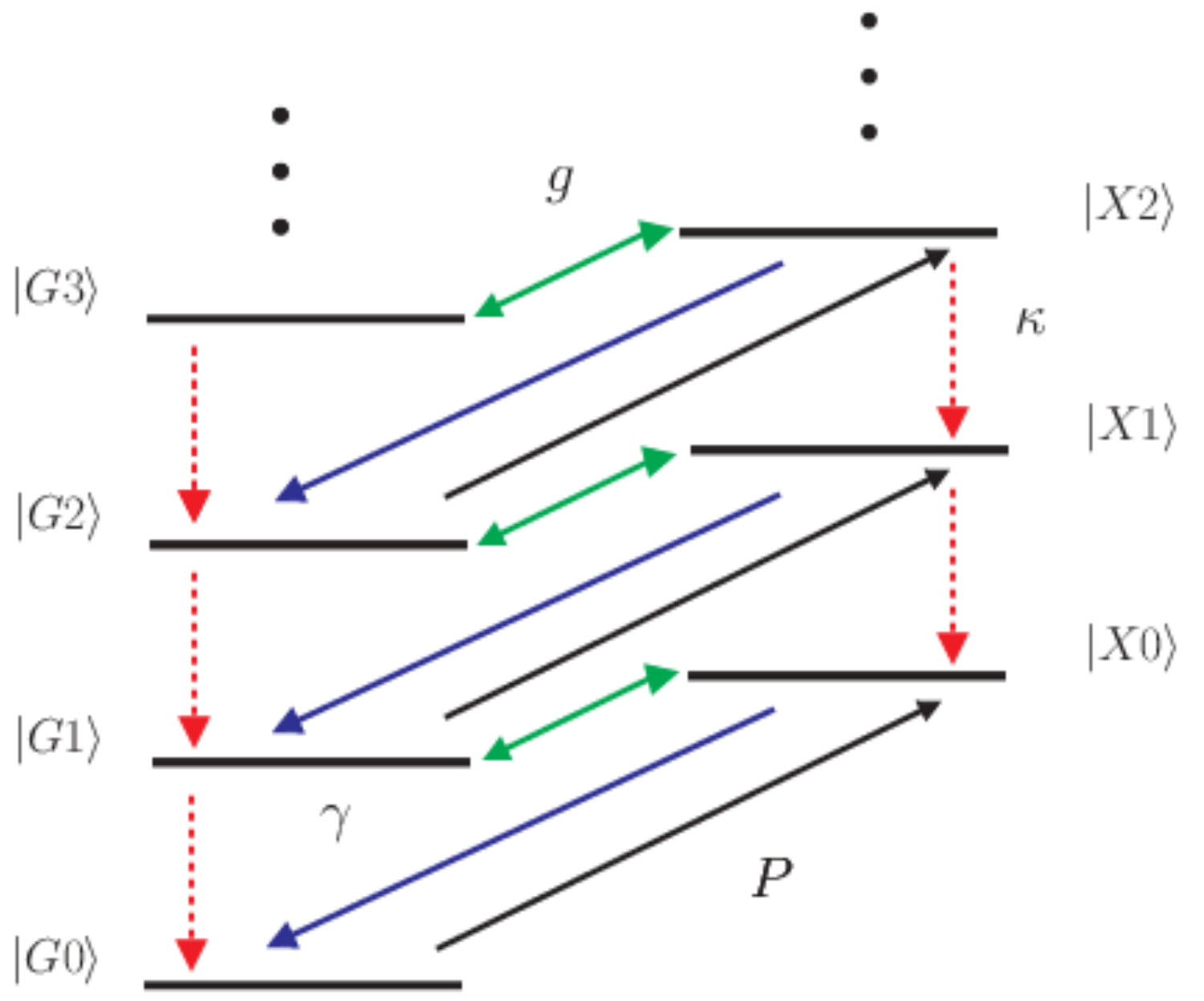}
\caption{Ladder of bared states for a two level quantum dot coupled to a single cavity mode. The double headed green arrow depics the matter coupling constant $g$, dashed red lines the emission of the cavity mode $\kappa$, solid black lines the exciton pumping rate $P$ and solid blue lines the spontaneous emission rate $\gamma$.}\label{qs}
\end{figure}

In order to apply our proposed method for calculating the emission spectrum in QD-Cavity system, we will consider a simple but illustrative system composed of a quantum dot interacting with a confined mode of the electromagnetic field inside a semiconductor cavity. This quantum system is well described by the Jaynes-Cummings Hamiltonian \cite{Cummings:1963}
\begin{equation}\label{eq:Hsystem}
\hat{H}_{S}=\omega_X\hat{\sigma}^{\dagger}\hat{\sigma}+(\omega_X-\Delta)\hat{a}^{\dagger}\hat{a}+g(\hat{\sigma}\hat{a}^{\dagger}+\hat{a}\hat{\sigma}^{\dagger}),
\end{equation}
where the quantum dot is described as a fermionic system with only two possible states, $\ket{G}$ and $\ket{X}$ are the ground and excited state, respectively. $\hat{\sigma}=|G\rangle\langle X|$ and $\hat{a}$ ($\hat{\sigma}^{\dagger}=|X\rangle\langle G|$ and $\hat{a}^\dagger$) are the annihilation (creation) operators for the fermionic system and the cavity mode. $g$ is the light-matter coupling constant, and we have set $\hbar=1$. We also define the detuning between frequencies  of the quantum dot and the cavity mode as $\Delta=\omega_X-\omega_a$, moreover $\omega_X$ is the energy to create an exciton and $\omega_a$ is the energy associated to the photons inside de cavity, respectively. This Hamiltonian system is far away for describing any real physical situation since it is completely integrable \cite{Scully:1996} and no measurements could be done since the light remains always inside the cavity.\\
In order to include the effects of environment on the dynamics of the system, we consider the usual approach to model an open quantum system by considering a whole system-reservoir hamiltonian which is frequently splitted in three parts. The first part corresponds to the system of quantum dot-microcavity. The second part is the hamiltonian of the reservoirs and finally, the third part which is a bilinear coupling between the system and the reservoirs \cite{Perea:2004}. After tracing out the degrees of freedom of all the reservoirs and assuming the validity of the Born-Markov approximation, one arrives to a master equation for the reduced density matrix of the system, 
\begin{eqnarray}\label{eq:master}
\frac{d\hat{\rho}_{S}}{dt}&=&i\left[\hat{\rho}_{S},\hat{H}_{S}\right]+\frac{\kappa}{2}(2 \hat{a} \hat{\rho}_{S} \hat{a}^{\dagger}-\hat{a}^{\dagger} \hat{a} \hat{\rho}_{S}-\hat{\rho}_{S} \hat{a}^{\dagger}\hat{a})\notag\\&+&\frac{\gamma}{2}(2 \hat{\sigma} \hat{\rho}_{S} \hat{\sigma}^{\dagger}-\hat{\sigma}^{\dagger} \hat{\sigma} \hat{\rho}_{S}-\hat{\rho}_{S} \hat{\sigma}^{\dagger}\hat{\sigma})\notag \\ 
&+&\frac{P}{2}(2 \hat{\sigma}^{\dagger} \hat{\rho}_{S} \hat{\sigma}-\hat{\sigma} \hat{\sigma}^{\dagger} \hat{\rho}_{S}-\hat{\rho}_{S} \hat{\sigma}\hat{\sigma}^{\dagger}).
\end{eqnarray}
Where $\gamma$ is the decay rate due to the spontaneous emission, $\kappa$ is the decay rate of the cavity photons across the cavity mirrors, and $P$ is the rate at which the excitons are being pumped. Fig.~\ref{systemscheme} shows a scheme of the simplified model of the QD-cavity system showing the processes of continuous pumping $P$ and cavity loses $\kappa$. The physical process begin when the light from the pumping laser enters into the cavity and excites one of the quantum dots in the QD layer. Thus, light from this source couples to the cavity and a fraction of photons escapes through the partly transparent mirror from the cavity and goes to the spectrometer for measurements of the emission of spectrum.\\ A general approach for solving the dynamics of the coupled system, consist of writting the Bloch equations for the reduced density matrix of the system in the bared basis. It is, an extended Hilbert space formed by taking the tensor product of the state vectors for each of the system components, ${\left\lbrace|G\rangle,|X\rangle\right\rbrace}\otimes \{{|n\rangle}\}^\infty_{n=0}$. In this basis, the reduced density matrix $\hat{\rho}_{S}$ can be written in terms of its matrix elements as $\rho_{S \alpha n,\beta m}=\langle\alpha n|\hat{\rho}_{S} |\beta m\rangle$. Hence, the Eq.~\eqref{eq:master} explicitly reads, 
\begin{eqnarray}\label{eq:maestra}
\frac{d\rho_{S \alpha n,\beta m}}{d\tau}&=&i\Big[(\omega_X-\Delta)(m-n)\rho_{S \alpha n,\beta m}\notag\\&+&\omega_X(\delta_{\beta X}\rho_{S \alpha n,Xm}-\delta_{\alpha X}\rho_{S Xn,\beta m})\Big]\notag\\&+&ig\Big[\Big(\sqrt{m+1}\delta_{\beta X}\rho_{S \alpha n,G m+1}\notag\\&+&\sqrt{m}\delta_{\beta G}\rho_{S \alpha n,X m-1}\Big)\notag\\&-&\Big(\sqrt{n}\delta_{\alpha G}\rho_{S X n-1,\beta m}\notag\\&+&\sqrt{n+1}\delta_{\alpha X}\rho_{S G n+1,\beta m}\Big)\Big]
\notag\\&+&\frac{\kappa}{2}\Big(2\sqrt{(m+1)(n+1)}\rho_{S \alpha n+1,\beta m+1}\notag \\ &-&(n+m)\rho_{S \alpha n,\beta m}\Big)
-\frac{\gamma}{2}\Big(\delta_{\alpha X}\rho_{S X n,\beta m}\notag \\&-&2\delta_{\alpha G}\delta_{\beta G}\rho_{S X n,X m}+\delta_{\beta X}\rho_{S \alpha n,X m}\Big)\notag\\&+&\frac{P}{2}\Big(2\delta_{\alpha X}\delta_{\beta X}\rho_{S Gn,Gm}-\delta_{\alpha G}\rho_{S Gn,\beta m}\notag\\&-&\delta_{\beta G}\rho_{S \alpha n,Gm}\Big).
\end{eqnarray}
Note that we use the convention that all indices written in greek alphabet are used for matter states and take values $\ket{G}$, $\ket{X}$, and the indices written in latin alphabet are used for Fock states and take values $0,1,2,3\dots$. Additionally, it is worth mentioning that our proposed method does not require to solve a system of coupled differential equations, instead of it, we solve a reduced set of algebraic equations that speed up the numerical solution. \\Prior to leaving this section, we point out that the number of excitations of the system is defined by the operator $\hat{N}=\hat{a}^{\dagger}\hat{a}+\hat{\sigma}^{\dagger}\hat{\sigma}$. The closed system and the number of excitations of the system is conserved, i.e., $[\hat{H}_S, \hat{N}]=0$. It allows us to organize the states of the system through the number of excitations criterion such that the density matrix elements $\rho_{Gn,Gn}$, $\rho_{Xn-1,Xn-1}$ $\rho_{Gn,Xn-1}$ and $\rho_{Xn-1,Gn}$ are related by having the same number of quanta. It 
is, subspaces of a fixed number of excitation evolve independently from each other. The Fig.~\ref{qs} shows a schematic representation of the action of the dissipative processes involved in the dynamics of the system according to the excitation number ($N_{exc}$).
\subsection{Emission spectrum of the cavity}\label{cavityspectrum}
\noindent
In order to compute the emission spectrum of the cavity, we will consider the two-time correlation function according to the Eq.~(\ref{eq-final}) for the photon operator as follows:
\begin{equation}
K(\tau)=\lim_{t\to\infty}\langle \hat{a}^{\dagger}(t+\tau)\hat{a}(t)\rangle.
\end{equation}
After performing the partial trace over the degrees of freedom of the system we have that
\begin{eqnarray}\label{ka}
K(\tau)&=&\sum_{\alpha,\beta,\gamma,l,m,n}\sqrt{(l+1)(m+1)}Tr_{R}[U_{\alpha l,\beta m}(\tau)\notag \\ &\times&\langle\beta m+1|\hat{\rho}^{(ss)}_{S\otimes R}|\gamma n\rangle U^{\dagger}_{\gamma n,\alpha l+1}(\tau)],
\end{eqnarray}
where the matrix elements for the time evolution operator are given by $U_{\alpha l,\beta m}(\tau)=\langle \alpha l|\hat{U}(\tau)|\beta m\rangle$ and $U^{\dagger}_{\gamma n,\alpha l+1}(\tau)=\langle \gamma n|\hat{U}^{\dagger}(\tau)|\alpha l+1\rangle$. In what follows, we assume the validity of the Markovian approximation, it means that the correlations between the system and the reservoir must be unimportant even at the steady state. Thus, the density operator system-reservoir can written as $\hat{\rho}^{(ss)}_{S\otimes R}=\hat{\rho}^{(ss)}_{S}\otimes\hat{\rho}^{(ss)}_{R}$ which implies that 
\begin{equation}\label{densitySR}
\langle\beta m+1|\hat{\rho}^{(ss)}_{S\otimes R}|\gamma n\rangle = \hat{\rho}^{(ss)}_{R}\langle\beta m+1|\hat{\rho}^{(ss)}_{S}|\gamma n\rangle.
\end{equation}
Replacing the previous expression in Eq.~(\ref{ka}), it is straightforward to shows that the two-time correlation function reads
\begin{equation}\label{corr}
 K(\tau)=\sum_{\alpha l}\sqrt{l+1}\langle\alpha l|\hat{G}(\tau)|\alpha l+1\rangle,
\end{equation}
where the Green's functions operator $\hat{G}(\tau)$ is given by
\begin{eqnarray}\label{op.green}
\hat{G}(\tau)&=&Tr_{R}\Big[\hat{U}(\tau)\hat{\rho}^{(ss)}_R\sum_{\beta \gamma m n}\Big(\sqrt{m+1}\ket{\beta m}\bra{\gamma n}\notag\\ 
&\times&\bra{\beta m+1}\hat{\rho}^{(ss)}_S\ket{\gamma n}\Big)\hat{U}^{\dagger}(\tau)\Big].
\end{eqnarray}
As we pointed out in section \ref{sec:QRT}, this operator must obey the same master equation as the reduced density operator of the system. In fact, the terms that only contribute in the Eq.~(\ref{corr}) are given by the matrix elements $G_{\beta m, \gamma n}(\tau)\equiv\bra{\beta m}\hat{G}(\tau)\ket{\gamma n}$ of the Green's functions operator. This is due to the fact that the projection operator $\ket{\beta m}\bra{\gamma n}$ enter into $\hat{G}(\tau)$ in the same way as into the reduced density operator of the system.\\ 
In order to identify these matrix elements, it should consider that for the QD-Cavity system the dynamics of the all coherences asymptotically vanish and remains only the reduced density matrix elements which are ruled by the number of excitations criterion, i.e. $\rho_{Gn,Gn}$, $\rho_{Xn-1,Xn-1}$, $\rho_{Gn,Xn-1}$, $\rho_{Xn-1,Gn}$. Then, the Eq.~(\ref{densitySR}) can be written as follows,
\begin{eqnarray} \label{diagonalrho}
\langle\beta m+1|\hat{\rho}^{(ss)}_{S\otimes R}|\gamma n\rangle &=& \hat{\rho}^{(ss)}_{R}\Big(\delta_{\beta G}\delta_{\gamma G}\delta_{m+1,n}\notag \\
&+& \delta_{\beta X}\delta_{\gamma X}\delta_{m,n-1}+\delta_{\beta G}\delta_{\gamma X}\delta_{m,n}\notag \\
&+& \delta_{\beta X}\delta_{\gamma G}\delta_{m+1,n-1}\Big)\rho^{(ss)}_{S\beta m+1,\gamma n}.\notag\\
\end{eqnarray}
By replacing the Eq.~(\ref{diagonalrho}) into Eq.~(\ref{op.green}) we find that the Green's functions operator explicitly reads
\begin{eqnarray}
\hat{G}(\tau)&=&Tr_R\Big[\hat{U}(\tau)\hat{\rho}^{(ss)}_R\sum_{m}\sqrt{m+1}\Big(\ket{G m}\bra{Gm+1}\notag \\
&\times& \rho^{(ss)}_{S Gm+1,Gm+1}+\ket{X m}\bra{X m+1}\rho^{(ss)}_{S X m+1,X m+1}\notag \\&+&\ket{G m}\bra{X m}\rho^{(ss)}_{S Gm+1,Xm}+\ket{X m}\bra{G m+2}\notag \\&\times&\rho^{(ss)}_{S X m+1,Gm+2}\Big)\hat{U}^{\dagger}(\tau)\Big].
\end{eqnarray}
Note that from this expression is easy to identify the nonzero matrix elements of the Green's functions operator that contribute to the emission spectrum. Finally, after performing the Laplace transform we have that the emission spectrum of the cavity is given by
\begin{eqnarray}
S(\omega)&=&\frac{1}{\pi n_c}\sum_l\sqrt{l+1}\Big(\tilde{G}_{Gl,Gl+1}(i\omega)\notag\\
         &+&\tilde{G}_{Xl,Xl+1}(i\omega)+\tilde{G}_{Gl,Xl}(i\omega)\notag \\ 
         &+&\tilde{G}_{Xl,Gl+2}(i\omega)\Big).
\end{eqnarray}
It is worth mentioning that the initial conditions may be obtained by evaluating the Green's function operator at $\tau=0$, then using the fact that the time evolution operators become the identity and $Tr_R[\hat{\rho}^{(ss)}_R]=1$, we obtain a set of initial conditions given by
\begin{eqnarray}\label{greencondini}
\tilde{G}_{Gl,Gl+1}(0)&=&\sqrt{l+1}\rho^{(ss)}_{S Gl+1,Gl+1},\notag \\
\tilde{G}_{Xl,Xl+1}(0)&=&\sqrt{l+1}\rho^{(ss)}_{S Xl+1,Xl+1},\notag \\
\tilde{G}_{Gl,Xl}(0)&=&\sqrt{l+1}\rho^{(ss)}_{S Gl+1,Xl},\notag \\
\tilde{G}_{Xl,Gl+2}(0)&=&\sqrt{l+1}\rho^{(ss)}_{S Xl+1,Gl+2}.
\end{eqnarray}
Note that this set of initial conditions corresponds to the asymptotic solution of the Bloch equations for the reduced density matrix of the system.
\subsection{Emission spectrum of the quantum dot}
\noindent
In order to compute the emission spectrum of the quantum dot, we will consider the two-time correlation function given by Eq.~(\ref{eq-final}), but for the case of the matter operator:
\begin{equation}
K(\tau)=\lim_{t\to\infty}\langle \hat{\sigma}^{\dagger}(t+\tau)\hat{\sigma}(t)\rangle.\notag
\end{equation}
It is straightforward to show after performing the partial trace over the degrees of freedom of the system that the two-time correlation function reads
\begin{equation}\label{corr2}
K(\tau)=\sum_{\alpha l}\delta_{\alpha X}\langle G l|\hat{G}(\tau)|\alpha l\rangle,
\end{equation}
where the Green's functions operator $\hat{G}(\tau)$ is given by
\begin{eqnarray}\label{op.green2}
\hat{G}(\tau)&=&Tr_{R}\Big[\hat{U}(\tau)\sum_{\beta \gamma m n}\Big(\delta_{\beta X}\ket{G m}\bra{\gamma m}\notag\\ 
&\times&\bra{\beta m}\hat{\rho}^{(ss)}_{S\otimes R}\ket{\gamma n}\Big)\hat{U}^{\dagger}(\tau)\Big].
\end{eqnarray}
Assuming again the validity of the Markovian approximation and taking into account the number of excitations criterion, we have that the density operator system-reservoir can be written as:
\begin{eqnarray} \label{diagonalrho2}
\langle\beta m|\hat{\rho}^{(ss)}_{S\otimes R}|\gamma n\rangle &=& \hat{\rho}^{(ss)}_{R}\Big(\delta_{\beta G}\delta_{\gamma G}\delta_{m,n}\notag \\
&+& \delta_{\beta X}\delta_{\gamma X}\delta_{m,n}+\delta_{\beta G}\delta_{\gamma X}\delta_{m,n+1}\notag \\
&+& \delta_{\beta X}\delta_{\gamma G}\delta_{m,n-1}\Big)\rho^{(ss)}_{S\beta m,\gamma n}.\notag\\
\end{eqnarray}
By inserting the Eq.~(\ref{diagonalrho2}) into Eq.~(\ref{op.green2}) we find that the Green's functions operator explicitly reads
\begin{eqnarray}
\hat{G}(\tau)&=&Tr_R\Big[\hat{U}(\tau)\hat{\rho}^{(ss)}_R\sum_{m}\Big(\ket{G m}\bra{Xm}\rho^{(ss)}_{S Xm,Xm}\notag \\
&+&\ket{G m}\bra{G m+1}\rho^{(ss)}_{S X m,G m+1}\Big)\hat{U}^{\dagger}(\tau)\Big].
\end{eqnarray}
Analogously as in section~\ref{cavityspectrum}, we identify the nonzero matrix elements of the Green's functions operator that contribute to the emission spectrum and after performing the Laplace transform the emission spectrum of the quantum dot is given by
\begin{equation}
S(\omega)=\frac{1}{\pi n_{\sigma}}\sum_l\Big(\tilde{G}_{Gl,Xl}(i \omega)+\tilde{G}_{Gl,Gl+1}(i\omega)\Big).
\end{equation}
where $n_{\sigma}=\langle\hat{\sigma}^{\dagger}\hat{\sigma} \rangle$ is the normalizing factor at the steady-state.
Taking into account that the initial conditions are obtained by evaluating the Green's function operator at $\tau=0$, we have the time evolution operators become the identity and $Tr_R[\hat{\rho}^{(ss)}_R]=1$, thus, we obtain a set of initial conditions given by
\begin{eqnarray}
\tilde{G}_{Gl,Xl}(0)&=&\rho^{(ss)}_{S Xl,Xl},\notag \\
\tilde{G}_{Gl,Gl+1}(0)&=&\rho^{(ss)}_{S Xl,Gl+1},\notag \\
\tilde{G}_{Xl,Xl+1}(0)&=&0,\notag \\
\tilde{G}_{Gl+2,Xl}(0)&=&0.
\end{eqnarray}
\section{Results and Discussion}\label{results}
\begin{figure}[h!]
\centering
\includegraphics[scale=.44]{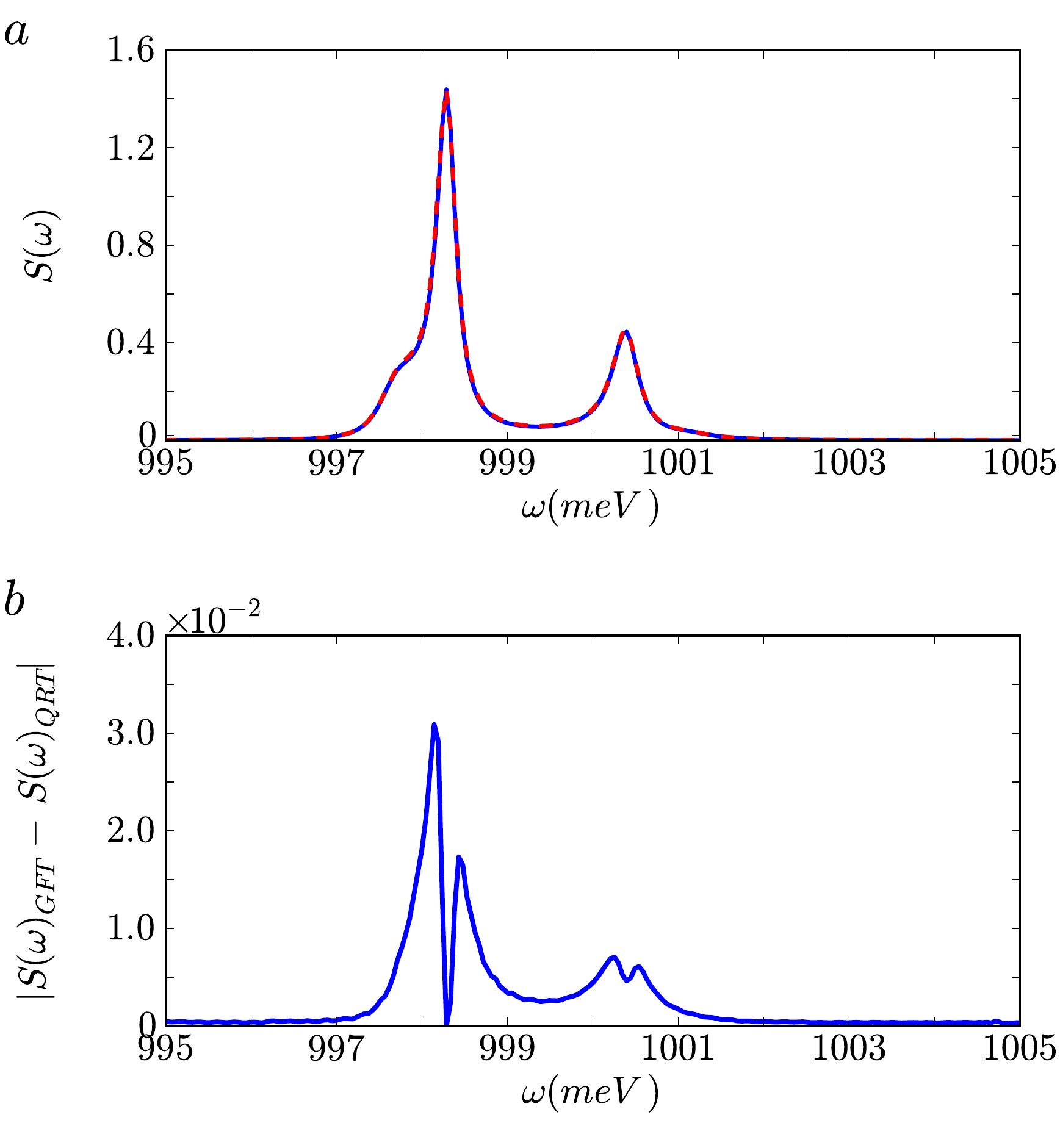}
\caption{Panel (a) shows a comparison of the emission spectrum of the cavity. The numerical calculation based on the Green's functions technique (GFT) is shown as solid blue line and the corresponding numerical calculation based on quantum regression theorem (QRT) approach is shown as dashed red line. Panel (b) shows in solid blue line the quantity $\vert S(\omega)_{GFT}-S(\omega)_{QRT}\vert$ as a measure of the difference in the numerical calculations of the emission spectrum of the cavity between two methods. 
}\label{panel2}
\end{figure}

\begin{figure}[h]
\centering
\includegraphics[scale=.44]{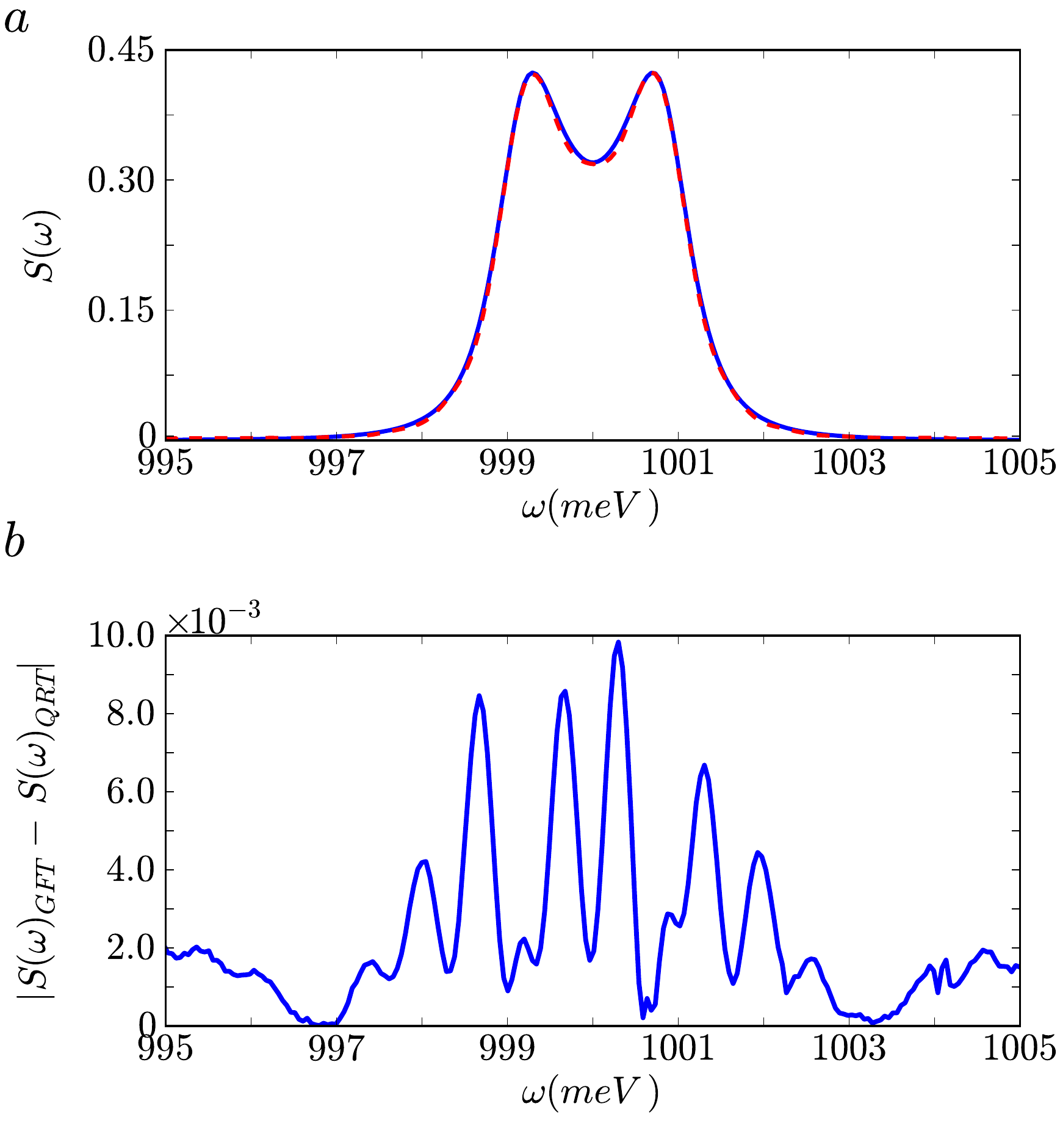}
\caption{Panel (a) shows a comparison of the emission spectrum of the cavity. The numerical calculation based on the Green's functions technique (GFT) is shown as solid blue line and the corresponding numerical calculation based on quantum regression theorem (QRT) approach is shown as dashed red line. Panel (b) shows in solid blue line the quantity $\vert S(\omega)_{GFT}-S(\omega)_{QRT}\vert$ as a measure of the difference in the numerical calculations of the emission spectrum of the cavity between two methods.
}\label{panel1}
\end{figure}

\begin{figure}[h!]
\centering
\includegraphics[scale=.44]{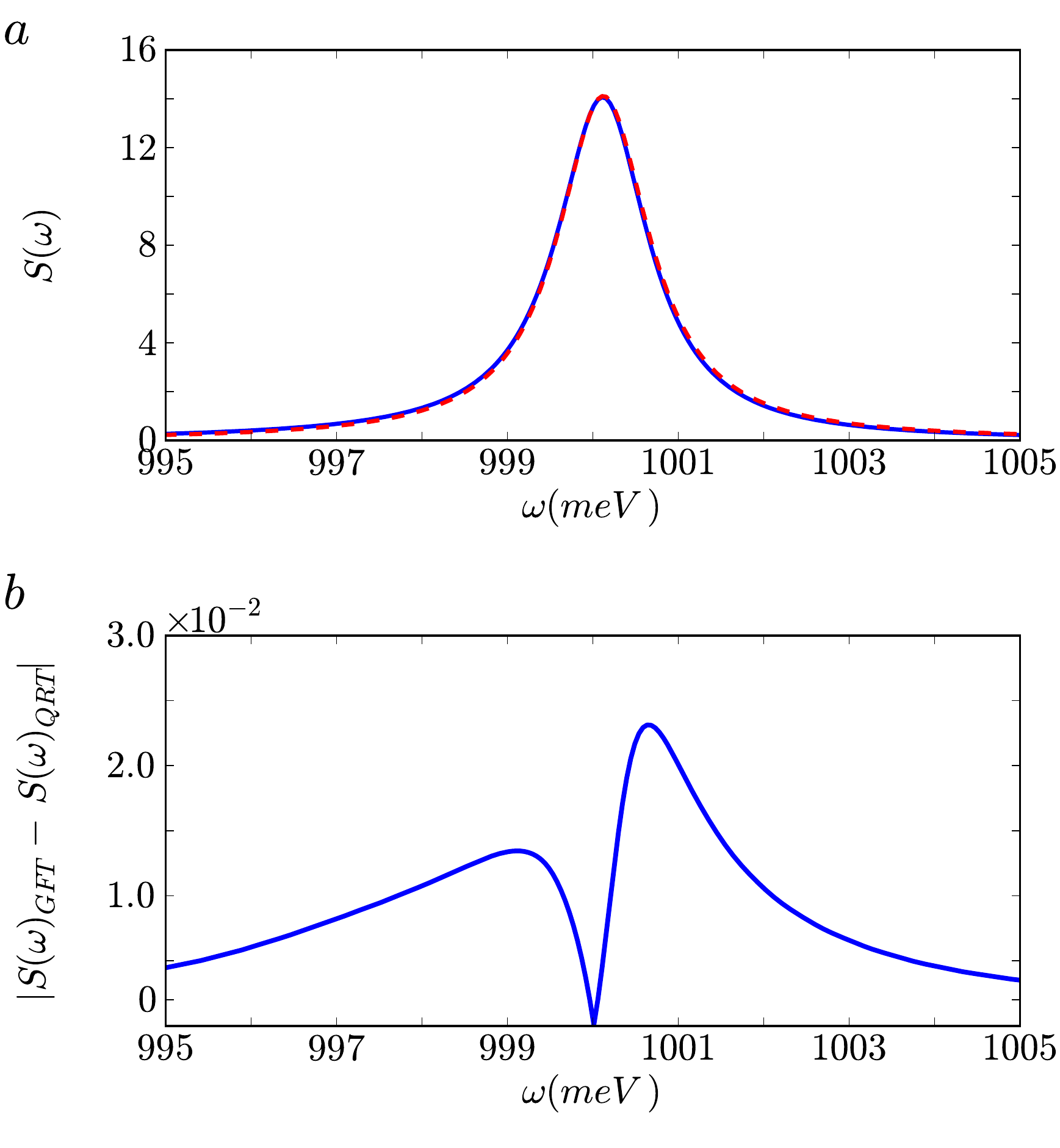}
\caption{Panel (a) shows a comparison of the emission spectrum of the quantum dot. See text and caption of Fig.~\ref{panel1} for details.}\label{panel3}
\end{figure}

In this section, we compare the numerical calculations based on GFT and QRT approach for the emission spectrum of the cavity as well as the quantum dot. Due to that the QD-Cavity system can display two different dynamical regimes by changing the values of the free parameters of the system and transitions between these two regimes can be achieved when the loss and pump rates are modified. Particularly, in the strong coupling regime the relation $P/\kappa\ll g$ holds and the relation $P/\kappa\gg g$ remains valid for the weak coupling regime. Fig.~\ref{panel2} shows the numerical calculations of the emission spectrum due to the cavity in the weak coupling regime, the parameters values are $g=1\,meV$, $\gamma=0.005\,meV$, $\kappa=0.2\,meV$, $P=0.3\,meV$, $\Delta=2\,meV$, $\omega_a=1000\,meV$. Panel (a) shows the emission spectrum for the GFT compared to the QRT approach. Panel (b) shows the quantity $\vert S(\omega)_{GFT}-S(\omega)_{QRT}\vert$ as a measurement of the error between the numerical calculations in the emission spectrum. For this set of parameters values, we can easily to identify two peaks associated to the modes of the cavity and the quantum dot, it is $\omega_a\approx998.3meV$ and $\omega_X\approx1000.3meV$, respectively. Fig.~\ref{panel1} shows the same calculations as in Fig.~\ref{panel2}, but in the strong coupling regime and the parameters values are $g=1\,meV$, $\gamma=0.005\,meV$, $\kappa=2\,meV$, $P=0.005\,meV$, $\Delta=0.0\,meV$, $\omega_a=1000\,meV$. In the case of resonance, the modes associated to the cavity and the quantum dot do not match, but
repel each other, resulting in a structure of two separate peaks a distance $2g\approx 2meV$. Fig.~\ref{panel3} shows the numerical calculations of the emission of spectrum due to the quantum dot with a high value of the rate $\kappa=5\,meV$ and a smaller, although non negligible, pumping $P=1\,meV$. The rest of parameters values are $g=1\,meV$, $\gamma=0.1\,meV$, $\Delta=5\,meV$, $\omega_a=1000\,meV$.\\
We observed that our numerical method based on GFT is in full agreement with the QRT approach and reproduces very well the spectrum of emission associated with this system. The quantity $\vert S(\omega)_{GFT}-S(\omega)_{QRT}\vert$ shows the discrepancy between both methods which is the order of $10^{-3}-10^{-2}$ as it is seen in Fig.~\ref{panel1} and Fig.~\ref{panel2}.  Mainly, the discrepancy between both methods is due to the numerical errors accumulated in the numerical integration of the Bloch equations in the QRT approach, it causes some differences in the spectrum with respect to the results computed by the GFT. Note that, there is no integration of any equations in the GFT, therefore, we expect a more accurate spectrum of emission. We mention that for the numerical calculations based on QRT approach, we have followed the Ref.~\cite{Vera:2009}. 
In order to test the performance of the numerical method, we regard four calculation times for computing the emission spectrum of the cavity based on GFT and QRT approach at different excitation numbers. Table~\ref{table01} shows in first column the excitation number, i.e. the truncation level in the bare-state basis for the numerical calculations involved. Second and third column show the results of elapsed time in seconds for both the GFT and QRT approach, respectively. Note that for comparison purposes all numerical calculations were performed at the same truncation level i.e. $N_{exc}=10$. Additionally, we have solved numerically the Bloch equations (see Eq.~(\ref{eq:maestra})) until time $t_{max}=2^{17}$ in order to obtain a good resolution in the frequency domain for the QRT approach, i.e. $\Delta\omega\approx0.048$. Hence, we have evaluated the emission spectrum for the GFT in a grid with the same resolution in the frequency domain (we emphasize that QRT approach is time consuming due to the number of coupled differential equations to be solved, rather than the number of evaluations in the grid size used). In addition, the numerical calculations were carried out with the same parameters values as in Fig.~\ref{panel1} for both GFT and QRT approach. We found that our numerical approach based on GFT is very efficient and accurate for calculating the emission spectrum in QD-Cavity systems. Moreover, this method can easily be implemented in the numerical linear algebra packages as well as in any programming language.
\begin{table}[h]
\caption{\label{table01} Comparison of calculation times between the Green's Functions Technique (GFT) and the Quantum Regression Theorem (QRT) in the numerical calculation of the emission spectrum of the cavity. The calculations were made using a commercial Intel(R) Core(TM) $i7-4770$ processor of $3.4$\,GHz $\times8$, and $12$ GB RAM.}
\begin{center}
\begin{tabular}{cccc}
\hline\hline
Excitation & Calculation times & Calculation times &  \\ 
 number  & for the GFT (s)& for the QRT (s) &  \\   
\hline   
$5$ &$\,\,\,\,\,\,\,0.4$ & $\,\,\,\,\,\,\,92.5$ & \\
$10$ &$\,\,\,\,\,\,\,2.0$ & $\,\,\,\,273.4$ & \\
$20$ &$\,\,\,\,14.0$ & $\,\,\,\,390.2$ & \\
$40$ &$\,100.2$ & $\,\,\,\,\,673.8$ & \\
\hline \hline
\end{tabular}
\end{center}
\end{table}  
\section {Conclusions}\label{sec:conclusions}
We have developed the Green's function technique as an alternative methodology to the QRT for calculating the two-time correlation functions in open quantum systems. In particular, we have shown the performance of the Green's function technique by calculating the emission spectrum in an open quantum system composed by a quantum dot embedded in a microcavity. This theoretical approach is rather general and allows to overcome the inherent theoretical difficulties presented in the direct application of the QRT, i.e., to find a closure condition on the set of operators involved in the dynamics equations, by considering that all coherences asymptotically vanish, and remains only the reduced density matrix elements which are ruled by the number of excitations criterion. We have shown that the Green's function technique offers several computational advantages, namely, the speeding up numerical computations via a transformation of the dynamics of the master equation in a set of linear algebraic equations, which are efficiently solvable by a numerical linear algebra routine, a faster convergence and significant reduction of computational time since the emission spectrum is calculated as a sum of terms of non-diagonal matrix elements of the reduced density operator of the system. We mention that our methodology can be extended for calculating the emission spectrum in significant situations of quantum dots in biexcitonic regime or involving coupled photonic cavities.
\section{Acknowledgements}
\noindent
This work was financed by Vicerrector\'ia de investigaciones of the Universidad del Quind\'io within the project with code $659$, and by Colciencias within the project with code $110156933525$, contract number $026-2013$, and HERMES code $17432$.

\end{document}